\documentclass[preprint,prl,aps]{revtex4-1}
\usepackage{amsmath}
\usepackage{harpoon}
\begin{document}
\date{\today}
\title{Correct $\Delta m^2_{ij}$ Dependence for Neutrino Oscillation Formulae}
\author{Randy A. Johnson}
\affiliation{University of Cincinnati, Cincinnati, OH 45221}

\begin{abstract}
The time translation operator for neutrino mass states is often taken to be $e^{-iEt/\hbar}$. This is not relativistically invariant.  In kaon mixing, physicists use $e^{-imc^2\tau/\hbar}$ where $\tau$ is the proper time of the kaon state.  The factor $mc^2\tau$ is the rest frame value of the four vector product $p_\mu x^\mu$ which is an invariant quantity.  If $-i p_\mu x^\mu$ is used in neutrino oscillation formulae instead of $-iEt$, the scale of the $\Delta m^2_{ij}$ is reduced by a factor of two.  

\end{abstract}
\maketitle

In the earliest papers on Dirac neutrino mixing \cite{Fritch, Cabibo}, the operator that has used to translate the neutrino mass states forward in time was taken to be $e^{-iEt/\hbar}$.  This has been used ever since ({\it e.g.}, \cite{LSND, SuperK, KARMEN, Kamland, Minos, DayaBay, Reno, DoubleChooz, MiniBooNE}).  In contrast for kaon mixing, the factor that is used to translate the neutral kaon mass states forward is $e^{-i m c^2 \tau / \hbar}$ where $\tau$ is the proper time of the kaon mass state.  The former phase factor is not relativistically invariant and, therefore, must be incorrect while the latter is proportional to the invariant product $p_\mu x^\mu$ evaluated in the particle's rest frame. It is ironic that Pontecorvio and collaborators, in their initial suggestion of Majorana neutrino oscillations ({\it e.g.}, \cite{Pontecorvo}), used the invariant time translation operator (although they never explicitly stated it).  Subsequent authors did not follow their lead.

Evaluating $p_\mu x^\mu$ in the laboratory frame gives:
\begin{align}
p_\mu x^\mu = \lbrace E,\overrightharp{p}c \rbrace \cdot \lbrace c t,\overrightharp{x} \rbrace & = E_{lab} c t_{lab} - p_{lab} c x_{lab} \nonumber\\
& = \left( \frac{E_{lab}}{\beta} - p_{lab} c\right) x_{lab} \nonumber\\
& = \left( {\frac{E^2_{lab}}{p_{lab} c}} - p_{lab} c\right) x_{lab} \nonumber\\
& = \frac{m^2 c^3}{p_{lab}} x_{lab} \nonumber
\end{align}
This gives a time translation operator of $e^{-i m^2 c^4 L/(E_{lab} \hbar c)}$ if the neutrino has traveled a distance $L$ and if its mass is sufficiently small such that $p_{lab} c \approx E_{lab}$.  The phase of the mass state actually changes twice as fast as with the operator usually used.  The corrected phase factor leads to the two neutrino oscillation formula
\begin{align}
P(\nu_\mu \rightarrow \nu_e ) & = \sin^2 2 \theta \sin^2 \left( \frac{\Delta m^2_{1,2}c^3 L} {2 E_{lab} \hbar} \right) \nonumber \\ 
& = \sin^2 2 \theta \sin^2 \left( 2.534 \frac{\Delta m^2_{1,2} L} { E_{lab}} \right) \nonumber
\end{align}
where, in the last equation, $\Delta m^2_{1,2}$ is in $eV^2$, $L$ is in $km$, and $E$ is in $GeV$. A similar change must be made to the many neutrino formulae.  While this correction does not invalidate any of the measurements of neutrino mass-squared differences, it does change the scale of those differences by a factor of two, $\Delta m^2_{actual}=\Delta m^2_{reported}/2$ for all neutrino oscillation measurements({\it e.g.}, \cite{PDG}).

\end{document}